# Edge reconstruction induces magnetic and metallic behavior in zigzag graphene nanoribbons

*Sudipta Dutta[1] and Swapan K. Pati*[1,2]

[1]Theoretical Sciences Unit and [2]New Chemistry Unit, Jawaharlal Nehru Centre for Advanced Scientific Research, Jakkur Campus, Bangalore-560064, India

ABSTRACT

The edge reconstruction of zigzag graphene nanoribbons to a stable line of alternatively fused seven and five membered rings with hydrogen passivation has been studied within density functional theory with both localized and extended basis approximations. Reconstruction of both edges results in a nonmagnetic metallic ground state, whereas the one edge reconstruction stabilizes the system in a ferromagnetic metallic ground state. The reconstructed edge suppresses the local spin density of atoms and contributes finite density of states at Fermi energy. Our study paves a new way to fabricate the metallic electrodes for semiconducting graphene devices with full control over the magnetic behavior without any lattice mismatch between leads and the channel.

Low dimensional carbon nanomaterials has been of great interest in recent times owing to their novel electronic, mechanical and optical properties with huge potential applications in this era of miniaturized electronic devices.[1,2] Recent discovery of graphene, a two dimensional monolayer of carbon atoms packed in a honeycomb lattice has enhanced the expectation and provided a new platform for experimental realization of many fascinating low dimensional properties which had been only theoretical artifacts over decades.[3,4]

Finite termination of graphene with smooth edges produces a quasi one-dimensional nano-structure, known as graphene nanoribbons (GNRs). A numerous theoretical studies on pure and modified GNRs show exciting magnetic, optical and spin filter properties with advanced device applications.[5-6] Experimental sophistications have made the understanding of this class of materials possible by cutting mechanically exfoliated graphene and lithographic patterning[7] or controlled epitaxial growth[8] or crystallographic etching by thermally activated metal nanoparticles[9] or following the synthetic chemistry route starting from small molecular building blocks.[10] A recent experimental technique based on solution dispersion and sonication showed the way to obtain narrow GNRs with smooth edges.[11] Unzipping carbon nanotubes has also been proven to be a good synthetic route in this regard.[12] Different geometrical terminations of two-dimensional graphene lattice lead to two edge geometries, namely, zigzag and armchair with largely varying properties originating from different boundary conditions. The zigzag GNRs (ZGNRs), in particular, has gotten tremendous attention due to its interesting magnetic and spintronic properties owing to its edge geometry.

Theoretically it has been observed that, the bare zigzag edge undergoes reconstruction where two consecutive hexagons transform into a pentagon and a heptagon, fused with each other to construct poly-azulene like edge.[13-15] This can be viewed as a line of Stone-Wales defect, which is energetically favorable by 0.35 eV/Å along the ribbon edge[13] with enhanced mechanical stability.[15]. Recent experimental findings confirm the stability of this reconstructed edge geometry.[16] However, it is surprising that, the existing literature rules out the possibility of hydrogen passivation of this reconstructed edge.[13,14] These results rather suggest that, the dangling bonds would form triple bonded dimers along the ribbon edge with bond angle 141.2° (see Figure S1). The proposed geometry would thus make the edge highly strained, since sp-hybridization would not allow such a large deviation in bond angle from its characteristic value of 180°. Instead, it is expected that the triple bonded edges would prefer to be hydrogenated to form $sp^2$-hybridized network with a bond angle near to 128.57° (characteristic for the heptagons). Moreover, the less stability of the triple bonds compared to double bonds (as evident from the higher heat of combustion of acetylene compared to ethylene) makes the hydrogenation more favorable. It is evident from the higher stability (0.7908 eV/Å along the edge) of the hydrogen passivated edges as compared to bare edge (see Supporting Information). This observation shows good agreement with the previous studies.[14] Interestingly, the hydrogen passivation of one reconstructed edge, leaving the other edge zigzag results in higher stability (0.9882 eV/Å, see the Supporting Information). In this article, we consider one edge and both edges reconstructed ZGNRs (Rc-ZGNRs) with hydrogen passivation and explore their structural, magnetic and electronic properties. We want to emphasize that, we are interested in the relative stability of the hydrogen passivated reconstructed edges compared to bare edges, in particular single edge reconstructed geometry, that has been ignored in all the previous studies.[13,14] Here, we propose that, in absence of hydrogen environment, the zigzag edge spontaneously reconstructs into a line of alternatively fused five and seven membered rings which takes up hydrogen while exposed in hydrogen environment to gain stability. Moreover, here we explore the single edge reconstructed structure in details, which has been observed experimentally[13b] without proper theoretical insight.

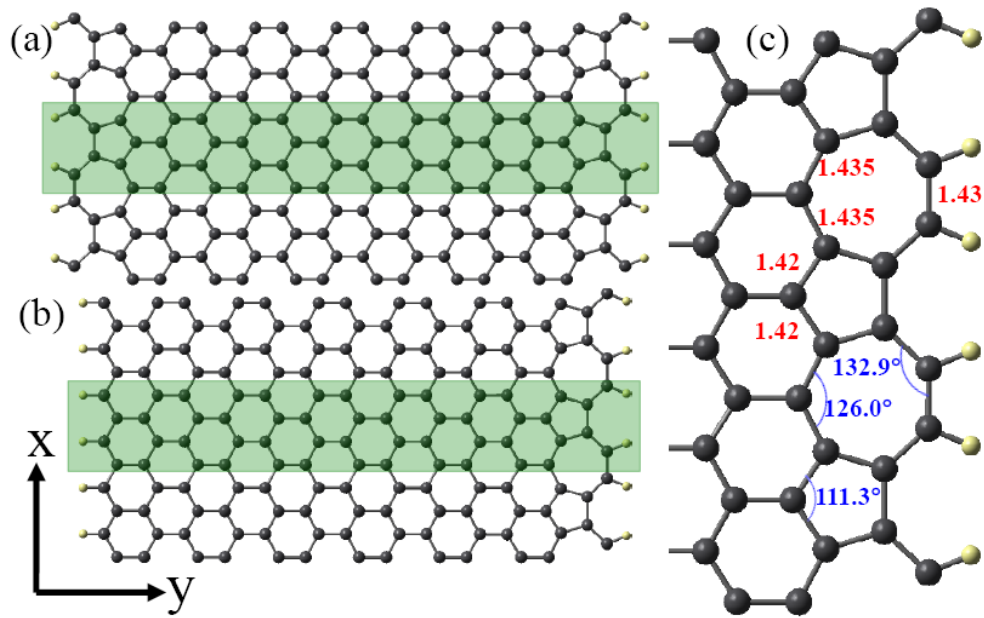

**Figure 1.** The structure of both edge (a) and one edge (b) reconstructed ZGNRs with hydrogen passivated edges. The highlighted regions show the unit cell with periodicity 4.9 Å along x axis, while the width is along y axis. (c) The edge geometry with selected bond lengths (Å) and bond angles.

We perform first principle periodic calculations based on density functional theory (DFT) as implemented in SIESTA package.[17] Spin polarized calculations have been performed within generalized gradient approximation (GGA), considering Perdew-Burke-Ernzerhof (PBE) exchange and correlation functionals[18] with double zeta polarized basis set. We consider quasi one-dimensional 8- and 12-ZGNRs (numbers specify the number of zigzag chains in the nanoribbon) with a poly-azulene edge in both sides and in one side, translated infinitely along the x axis, as shown in Figure 1(a) and (b) respectively. Lattice vectors along y and z axis have been considered as 50Å and 16Å respectively to create sufficient vacuum in order to avoid the interactions within adjacent unit cells. We consider 400 Ry energy cutoff for a real space mesh size and a k-point sampling of 36 k points, uniformly distributed along the one-dimensional Brillouin zone. We relax all the structures with initial guess of highest spin and lowest spin configurations. Here we present the results for ground state of edge reconstructed 12-ZGNRs. The results obtained from SIESTA with localized basis have been verified with extended plane augmented wave basis approximation as implemented within the DFT package, VASP.[19]

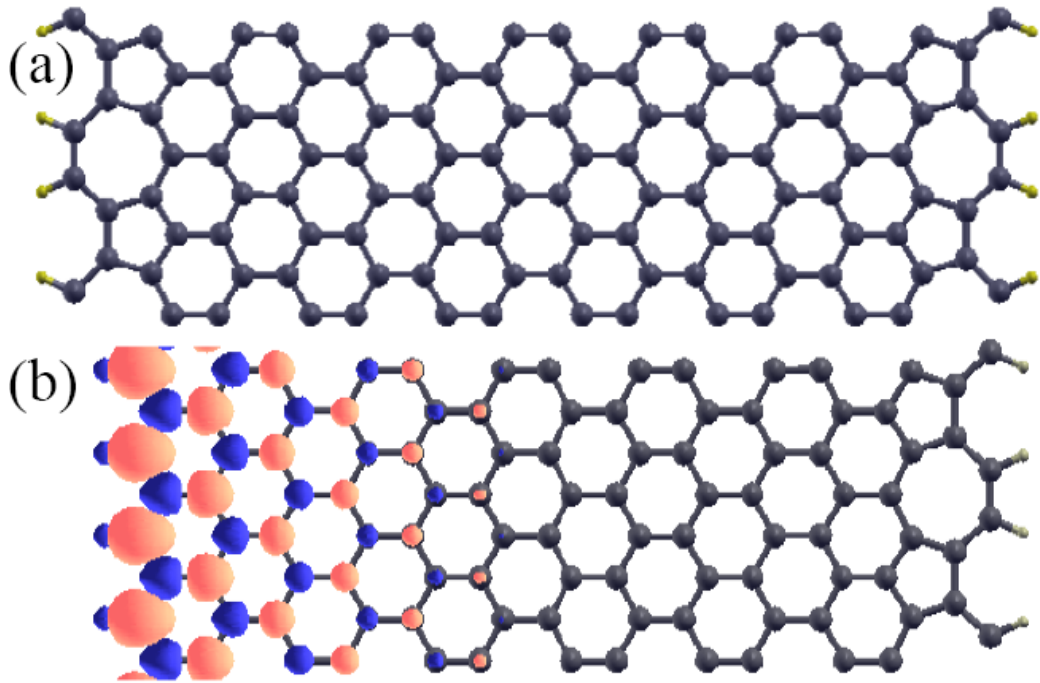

**Figure 2.** The spin density profile for (a) both edge and (b) one edge Rc-ZGNRs. In case of both edge Rc-ZGNR, the absence of spin density throughout the structure specifies the net zero local spin moment.

We find that, the both edge Rc-ZGNRs stabilize in nonmagnetic ground state as observed in previous studies[14] with slight elongation of the bonds constructing the heptagons as shown in Figure 1(c) and Figure S1. However, the structure maintains its $sp^2$-hybridized network throughout the ribbon lattice as can be seen from the relaxed bond lengths and bond angles. The reconstruction of edges suppresses the local spin moments (see Figure 2(a)), destroying the bipartite nature of opposite spin preferences in neighboring sites of pure ZGNRs.[6] However, in case of one edge Rc-ZGNRs, although the local spin moments are suppressed along the reconstructed edge, the abundance of one kind of spin along the unreconstructed zigzag edge (see Figure 2(b)) makes the ground state magnetic. The geometry optimization of corresponding super cells with four unit cells shows consistent behavior. Thus our study shows that, by selective reconstruction of ZGNRs, one can control the magnetic behavior.

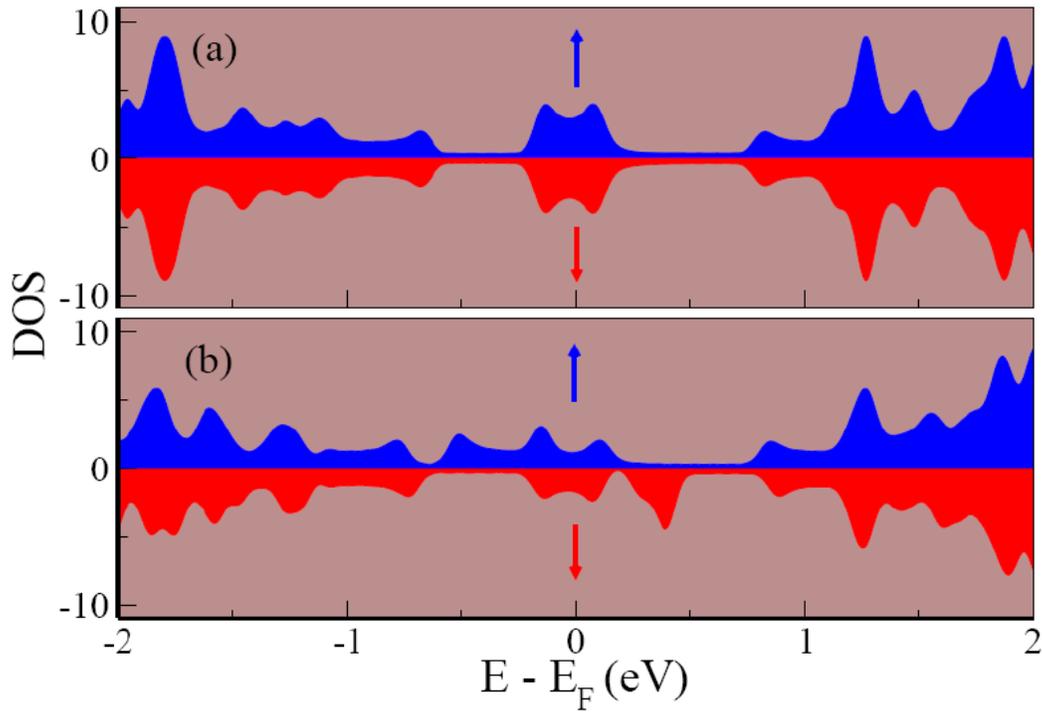

**Figure 3.** The density of states for (a) both edge and (b) one edge Rc-ZGNRs for up spin and down spin separately.

The presence of finite DOS at Fermi energy for both the Rc-ZGNRs (1(a) and 1(b)) results in metallic behavior (see Figure 3). It has been observed earlier that, the haeckelite structures consisting of equal number of pentagons and heptagons with any number of hexagons exhibit metallic behavior.[20] The DOS of both edge Rc-ZGNRs shows similar profile for both the spins (see Figure 3(a)) due to the geometrical symmetry in both the edges. However, the asymmetric edges in case of one edge Rc-ZGNRs make the DOS spin polarized as shown in Figure 3(b).

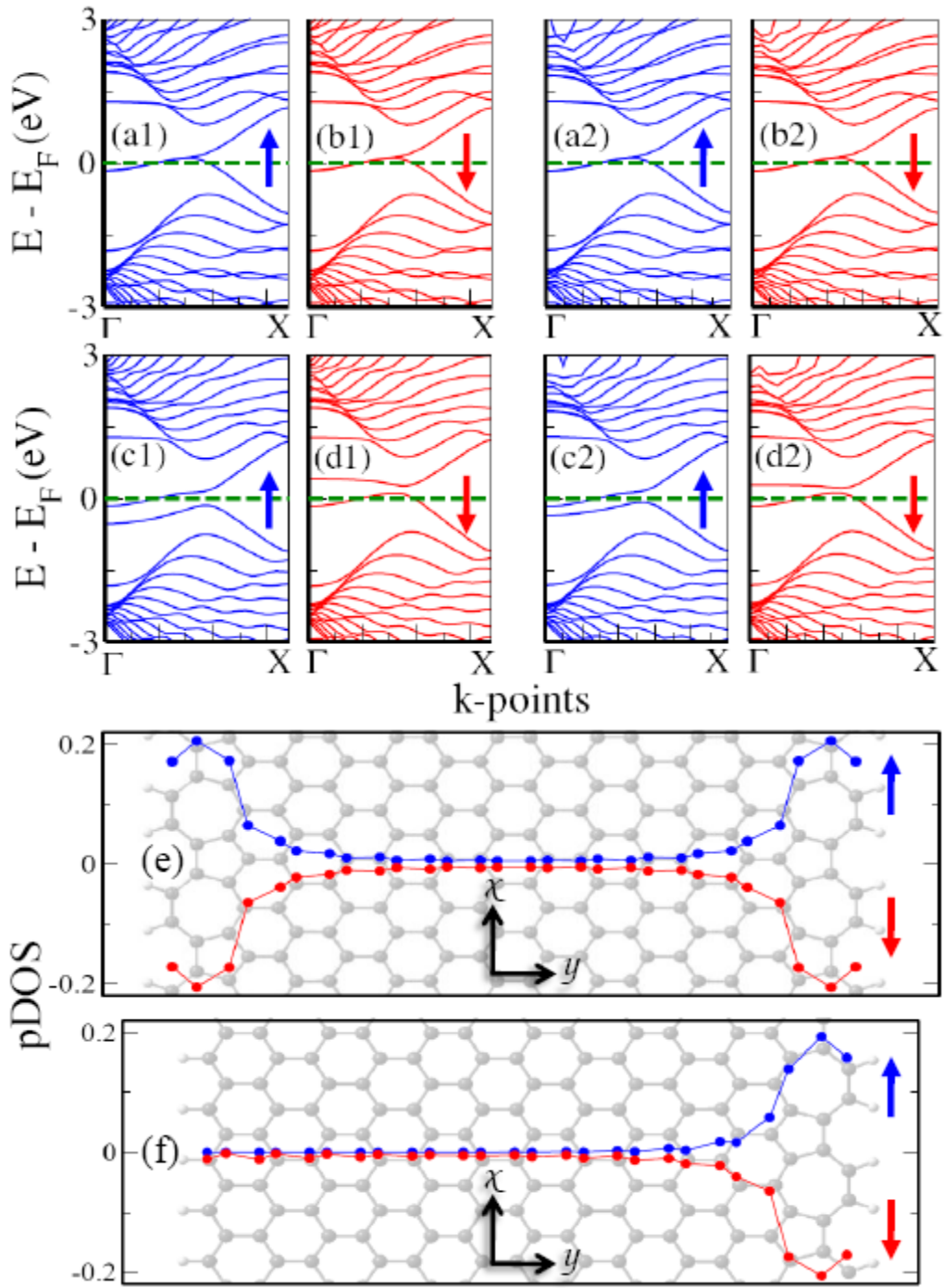

**Figure 4.** Band structures of both edge (a, b) and one edge (c, d) Rc-ZGNRs for up spin (a, c) and down spin (b, d). The results of SIESTA and VASP have been denoted as 1 and 2 respectively. (e) and (f) Shows the spin polarized average pDOS at Fermi energy for the atoms along the ribbon width according to their y axis coordinates as shown in background picture for both and one edge Rc-ZGNRs respectively.

The band structure analysis for up and down spins of both edge (Figure 4(a) and 4(b)) and one edge (Figure 4(c) and 4(d)) Rc-ZGNRs shows the presence of dispersive bands, characteristics of metallic behavior at the Fermi energy. The similar band structure profiles for up spin and down spin in case of both edge Rc-ZGNRs make the ground state nonmagnetic. However, the difference in up spin and down spin band structures leads to magnetic ground state in case of one edge Rc-ZGNRs. The excellent agreement of results from localized and extended basis approximations unambiguously confirms this induced metallicity in quasi 1-D systems. Our detail wave function analysis to understand the origin of metallicity suggests that, the $p_z$ orbitals of the carbon atoms at reconstructed edges form the dispersive metallic bands. However, at Brillouin zone boundary their contribution disappears. This can be further argued from the projected DOS (pDOS) analysis at Fermi energy of individual carbon atoms along the ribbon width. It clearly shows that the maximum contribution to the metallic DOS comes from the reconstructed edges (see Figure 4(e) and 4(f)). Thus, the semiconducting ZGNRs turn metallic upon single or both edge reconstruction. We also find that the metallicity sustains with even high external electric field along the cross ribbon width (see Figure S2). Although the up spin and down spin DOS in case of one edge Rc-ZGNRs behave differently in response to the external electric field, the metallic property remains unaffected. All these observations are consistent for Rc-ZGNRs of different widths.

In conclusion, we have studied the edge reconstruction of ZGNRs to a line of fused pentagons and heptagons with hydrogen passivated edges. Although the both edge Rc-ZGNRs are reported in some previous studies, our report on one edge Rc-ZGNRs and their interesting properties is first of its kind. The edge reconstruction of semiconducting ZGNRs at one and both edges provides higher stability, resulting in ferromagnetic and non-magnetic metals, respectively. Our study shows new inroads to fabricate the metallic electrodes for the semiconducting graphene devices with full control over its magnetic behavior without any lattice mismatch and motivates for further exploration for its applications in advanced devices.

**ACKNOWLEDGMENT** S.D. acknowledges CSIR, Govt. of India for research fellowship and S.K.P. acknowledges DST and CSIR, Govt. of India for research grant.

**SUPPORTING INFORMATION** The structure, geometry and the DOS for one and both edge Rc-ZGNRs in presence of external electric field.

# Edge reconstruction induces magnetic and metallic behavior in zigzag graphene nanoribbons

*Sudipta Dutta[1] and Swapan K. Pati\*[1,2]*

[1]Theoretical Sciences Unit and [2]New Chemistry Unit, Jawaharlal Nehru Centre for Advanced Scientific Research, Jakkur Campus, Bangalore-560064, India

**The structure, geometry and stability of edge reconstructed zigzag graphene nanoribbons in absence and in presence of hydrogen passivation:**

We have studied the zigzag graphene nanoribbons (ZGNRs) of different widths (with 8 and 12 zigzag chains) with edge reconstruction in one and both edges. The reconstructed edge consists of alternately fused seven and five membered rings to form a poly-azulene like edge geometry. We have studied the edge reconstructed ZGNRs (Rc-ZGNRs) both in absence and in presence of hydrogen passivation at the edges. The relaxed edge geometry for reconstructed one edge and both edges resembles each other.

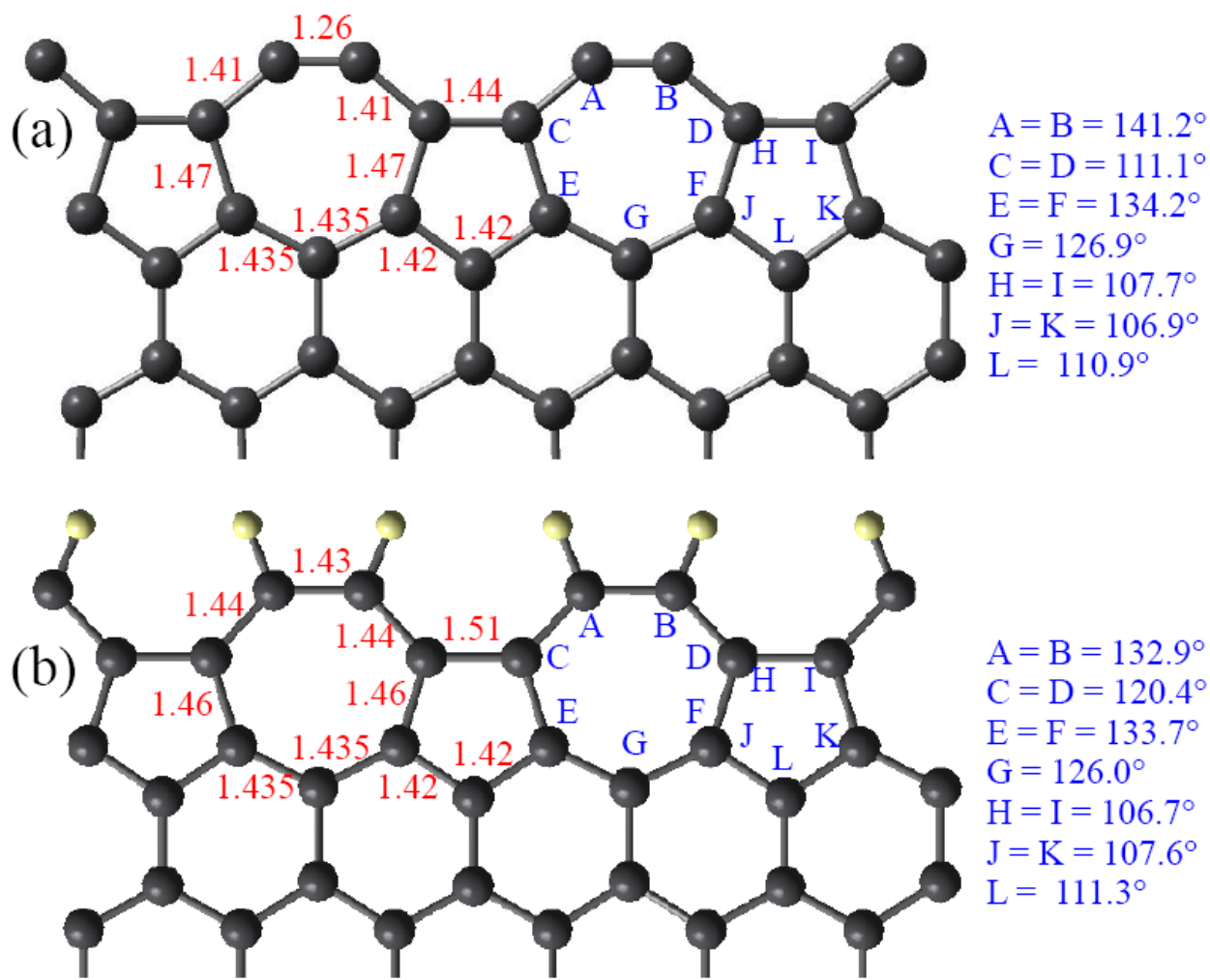

**Figure S1.** The edge geometry of the reconstructed edges of ZGNRs in absence (a) and in presence (b) of hydrogen passivation with selected bond lengths (Å) and bond angles.

It can be seen from Figure S1 that, the reconstructed edge without hydrogen passivation forms triple bonded dimers along the edges. The bond lengths along the edges (1.26 Å) resemble fairly well with the sp-hybridized carbon-carbon bond lengths in acetylene and suggests a sp-hybridization along the reconstructed edges. However, the bond angle (141.2°) at the reconstructed edge without hydrogen passivation deviates severely from the characteristic bond angle of the sp-hybridization (180°). Therefore this edge becomes highly strained. This observations show good agreement with the previous studies.[1]

The passivation of the reconstructed edge with hydrogen makes the edge sp$^2$-hybridized with its characteristics bond lengths and the bond angles also become comparable with the characteristic bond angles (128.57°) for seven membered rings. Thus the edge gains stability. To give a numerical estimation of this stabilization, we calculate the stabilization energy of the hydrogen passivated edges considering the following equation.

$$E_{stabilization} = E_{Rc-ZGNR+H} - E_{Rc-ZGNR} - \frac{N_H}{2} \times E_{H_2}$$

where, $E_{stabilization}$, $E_{Rc-ZGNR+H}$, $E_{Rc-ZGNR}$ and $E_{H_2}$ are the stabilization energy of the hydrogen passivated edge, total energy of system with hydrogen passivation, total energy of the system without hydrogen passivation and the energy of one hydrogen molecule, respectively. $N_H$ is the number of hydrogen atoms in the system.

For the 12-ZGNR with one reconstructed edges, the stabilization energy is,

$E_{stabilization} = 0.9882$ eV/Å along the ribbon edge

For the 12-ZGNR with both reconstructed edges, the stabilization energy is,

$E_{stabilization} = 0.7908$ eV/Å along the ribbon edge

These high stabilization energies of the hydrogen passivated edges compared to the bare edges suggests the preference of the edge hydrogenation and we pursue further calculations with only hydrogen passivated Rc-ZGNRs.

**Density of states (DOS) in presence of external electric field:**

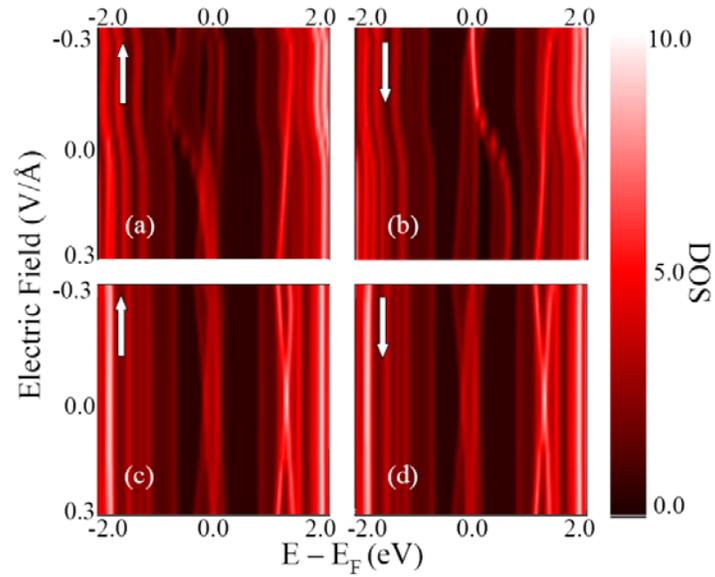

***Figure S2.*** Contour plot of the up spin (a) and down spin (b) density of states for one edge reconstructed 12-ZGNR as function of energy, scaled with respect to the Fermi energy and external applied electric field. (c) and (d) represents the same for up spin and down spins respectively for the both edge reconstructed 12-ZGNR. All the plots suggest that, the metallic behavior of the single and both edge reconstructed 12-ZGNRs sustains large external electric field.